\documentstyle[11pt,aasms4,epsfig]{article} 
 
\begin{document} 
\tighten 
\title{New non-Gaussian feature in COBE-DMR Four Year Maps}  
\author{Jo\~{a}o Magueijo} 
\affil{Theoretical Physics,  
Imperial College, Prince Consort Road, London SW7 2BZ, UK} 
 
\begin{abstract} 
We extend a previous bispectrum analysis of the Cosmic Microwave  
Background temperature anisotropy, allowing for the presence of  
correlations between different angular scales. We find a strong 
non-Gaussian signal in the ``inter-scale'' components 
of the bispectrum: 
their observed values concentrate close to zero instead of 
displaying the scatter expected from Gaussian maps. This signal 
is present over the range of multipoles $\ell=6 -18$, in 
contrast with previous detections. We attempt to attribute this effect 
to galactic foreground contamination, pixelization effects,  
possible anomalies in the noise, documented 
systematic errors studied by  
the COBE team, and the effect of assumptions used in our Monte Carlo 
simulations. Within this class of systematic errors 
the confidence level for rejecting Gaussianity varies 
between $97\%$ and $99.8\%$. 
\end{abstract} 
 
 
\section{Introduction} 
\label{introduction} 
In a recent {\it Letter} \cite{fmg} found strong evidence 
for non-Gaussianity 
in the anisotropy of the Cosmic Microwave Background (CMB) 
temperature.  
This detection was followed by similar claims from \cite{nfs98} and 
\cite{pvl98}, and caused considerable consternation among theorists 
(see \cite{nvs} for a discussion). These 
three groups employed very different statistical tools, but used the  
same dataset -- the COBE-DMR 
4 year maps. Recent work by \cite{bt99} confirmed these 
measurements,
but \cite{banday} cast 
doubts upon the cosmological origin of the observed signals. 
We clearly do not fully understand some of the less
conspicuous systematic 
errors associated with DMR maps. 
We feel that the origin of the observed
non-Gaussian features will probably not be conclusively identified 
before an independent all-sky dataset becomes available. 
 
In this Letter we revisit and complete the analysis of \cite{fmg}. 
In that work the possibility of departures from Gaussianity was 
examined in terms of the bispectrum. Given a full-sky map,  
$\frac{\Delta T}{T}({\bf n})$, this may be expanded into Spherical 
Harmonic functions: 
\begin{eqnarray} 
\frac{\Delta T}{T}({\bf n})=\sum_{\ell m}a_{\ell m}Y_{\ell m}({\bf n}) 
\label{almdef} 
\end{eqnarray} 
The coefficients $a_{\ell m}$ may then be combined into  
rotationally invariant multilinear forms (see \cite{santa} 
for a possible algorithm). The most general 
cubic invariant is the bispectrum,  and is given by 
\begin{eqnarray} 
{\hat B}_{\ell_1\ell_2\ell_3}&=&\alpha_{\ell_1\ell_2\ell_3} 
\sum_{m_1m_2m_3}\left  
( \begin{array}{ccc} \ell_1 & \ell_2 & \ell_3 \\ m_1 & m_2 & m_3 
\end{array} \right ) a_{\ell_1 m_1}a_{\ell_2 m_2} a_{\ell_3 m_3} 
\nonumber \\ 
\alpha_{\ell_1\ell_2\ell_3}&=&\frac{1}{(2\ell_1+1)^{\frac{1}{2}} 
(2\ell_2+1)^{\frac{1}{2}}(2\ell_3+1)^{\frac{1}{2}}}\left ( 
\begin{array}{ccc} \ell_1 
 & \ell_2 & \ell_3 \\ 0 & 0 & 0 
\end{array} \right )^{-1} 
\label{bispec} 
\end{eqnarray} 
where the $(\ldots)$ is the Wigner $3J$ symbol.  
In \cite{fmg} correlations between different multipoles were ignored, 
and so $\ell_1=\ell_2=\ell_3$. Here we consider bispectrum 
components sensitive to correlations between different scales. 
Selection rules require that $\ell_1+\ell_2+\ell_3$ be even. 
The simplest chain of correlators is therefore  
${\hat A_\ell}=B_{\ell-1\, \ell \,\ell+1}$, with $\ell$ even.  
Other components, involving more distant multipoles, may be considered 
but they are very likely to be dominated by noise; it is
natural to assume that possible non-Gaussian inter-scale
correlations decay with $\ell$ separation. 
We shall consider a ratio  
\begin{eqnarray} 
J^3_\ell &=& { {\hat A}_{\ell} 
\over ({\hat C}_{\ell-1})^{1/2}({\hat C}_{\ell})^{1/2} 
({\hat C}_{\ell+1})^{1/2}} 
 \label{defI} 
\end{eqnarray} 
where ${\hat C}_\ell=\frac{1}{2\ell+1}\sum_m|a_{\ell m}|^2$.  
This quantity is dimensionless, and is invariant under rotations 
and parity. It extends the $I^3_\ell$ statistic
used in \cite{fmg}~\footnote{Taking the modulus is not necessary 
to ensure 
parity invariance, contrary to the statement made in \cite{fmg}.
This does not affect any of the conclusions, since
$P(I^3_\ell)=P(-I^3_\ell)$ for a Gaussian process.}.

The theoretical importance of the bispectrum as a non-Gaussian 
qualifier has been  recognized in a number of publications
(\cite{luo94}, \cite{peebles1}, 
\cite{sg98}, \cite{gs98}, \cite{wang}). 
\cite{kog96a} measured the pseudocollapsed 
and equilateral three point function of the DMR four year data.
The bispectrum may be
regarded as the Fourier space counterpart of the three point function.
The work presented in this {\it Letter}, 
combined with \cite{fmg} and \cite{heav},
provides a complete set of signal dominated bispectrum components 
inferred from  DMR maps.

\section{The publicly released 4 year maps} 
We first consider the  inverse-noise-variance-weighted, 
average maps of  
the 53A, 53B, 90A and 90B {\it COBE}-DMR channels, with monopole 
and dipole removed, at resolution 6, in galactic and ecliptic  
pixelization. We use the   
extended galactic cut of \cite{banday97}, and  
\cite{benn96} to remove most of the emission from the 
plane of the Galaxy. 
To  estimate the $J^3_\ell$s we set 
the value of the pixels within the galactic cut to 0 and  
the monopole and dipole of the cut map to zero.  
We then integrate the 
map multiplied with spherical harmonics  
to obtain the estimates of the 
$a_{\ell m}$s and apply equations \ref{bispec} and \ref{defI}. 
The observed $J^3_\ell$ are to be compared with their distribution 
$P(J^3_\ell)$ as inferred from Monte Carlo simulations in which  
Gaussian maps are subject to DMR noise and galactic cut.
In simulating DMR noise we take into account the full
noise covariance matrix, as described in \cite{corn}. This includes
correlations between pixels $60^\circ$ degrees apart. 

\begin{figure} 
\centerline{\psfig{file=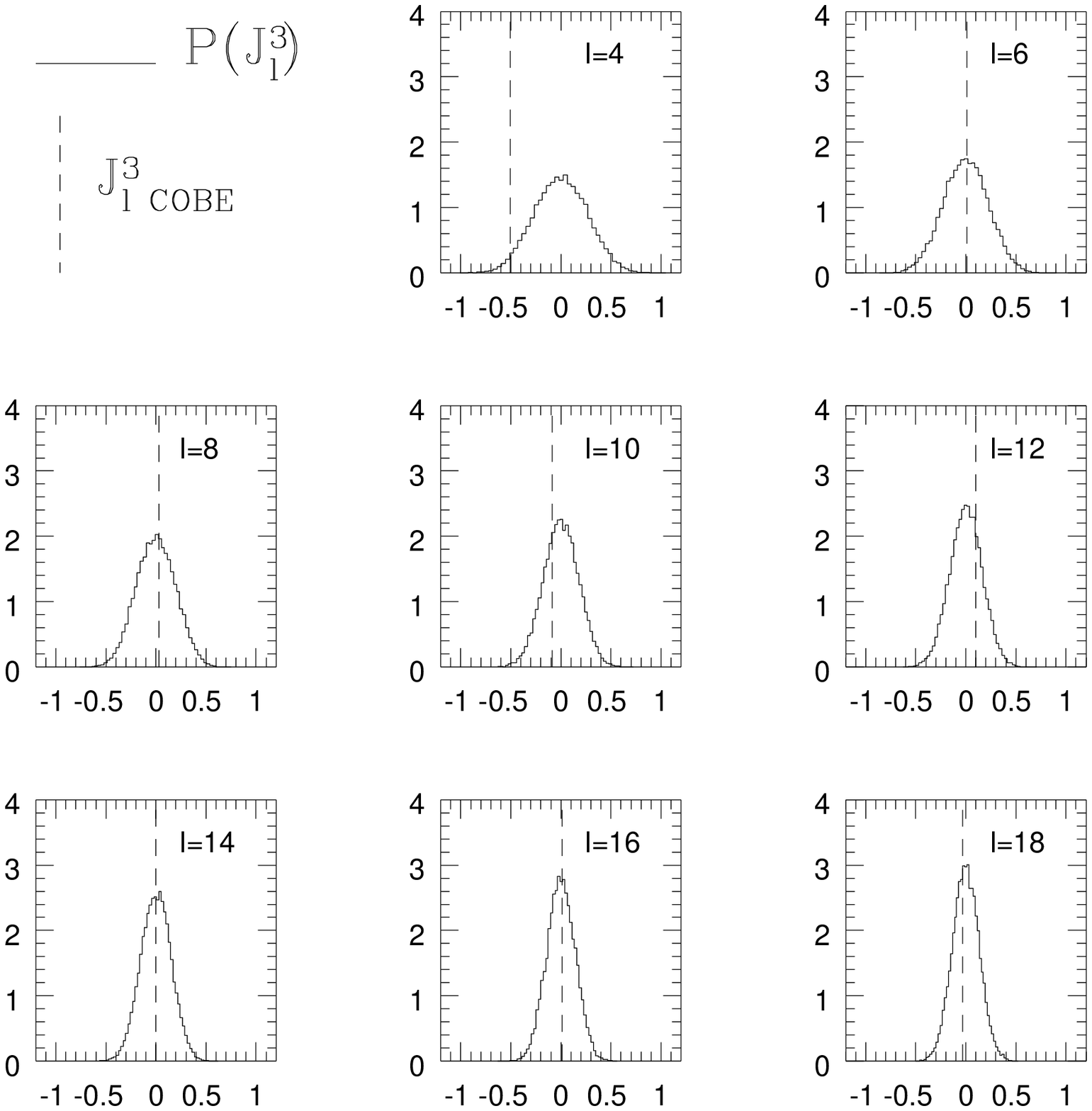,width=12cm}} 
\caption{The vertical thick dashed line represents the value  
of the observed 
$J^3_\ell$.  The solid line is the probability distribution function 
of $J^3_\ell$ for a Gaussian sky with extended galactic cut and 
DMR noise, as inferred from 25000 realizations. } 
\label{fig1} 
\end{figure}

The results are displayed in Fig.~\ref{fig1}. 
Leaving aside the deviant $J^3_4$, 
it is blatantly obvious
 that the observed $J^3_\ell$ do not exhibit the scatter 
around zero implied by $P(J^3_\ell)$ in the range of scales  
$\ell=6-18$. This may be mathematically formalized by means of a  
goodness of fit statistic, such as the ``chi squared'': 
\begin{equation}\label{presc} 
X^2={1\over N}{\sum_\ell X_\ell^2}= 
{1\over N}{\sum_\ell (-2\log P_\ell(J^3_\ell)  
+ \beta_\ell),} 
\end{equation} 
where the constants $\beta_\ell$ are defined so that for each term 
of the sum $\langle X_\ell^2\rangle=1$.  
As explained in \cite{fmg} this quantity reduces to the usual 
chi squared when the distributions $P$ are Gaussian. When $P$
is not Gaussian, $X^2$ goes to infinite where $P$ goes to zero, 
reaches a minimum (usually around zero) at the peak of $P$,
and has average 1. Hence $X^2$ does for a non-Gaussian $P$ what 
the usual chi squared does for a Gaussian $P$. A good 
fit is represented by $X^2\approx 1$. If $X^2\gg 1$ the data  
is plagued by deviants, that is observations far in the tail of the  
theoretical distribution.  
If $X^2\ll 1$ the observations fail to exhibit the scatter predicted 
by the theory, concentrating 
uncannily on the peak of the distribution. 
Both cases present grounds for rejecting the hypothesis embodied 
in $P(J^3_\ell)$, in our case Gaussianity.  
 
\begin{figure} 
\centerline{\psfig{file=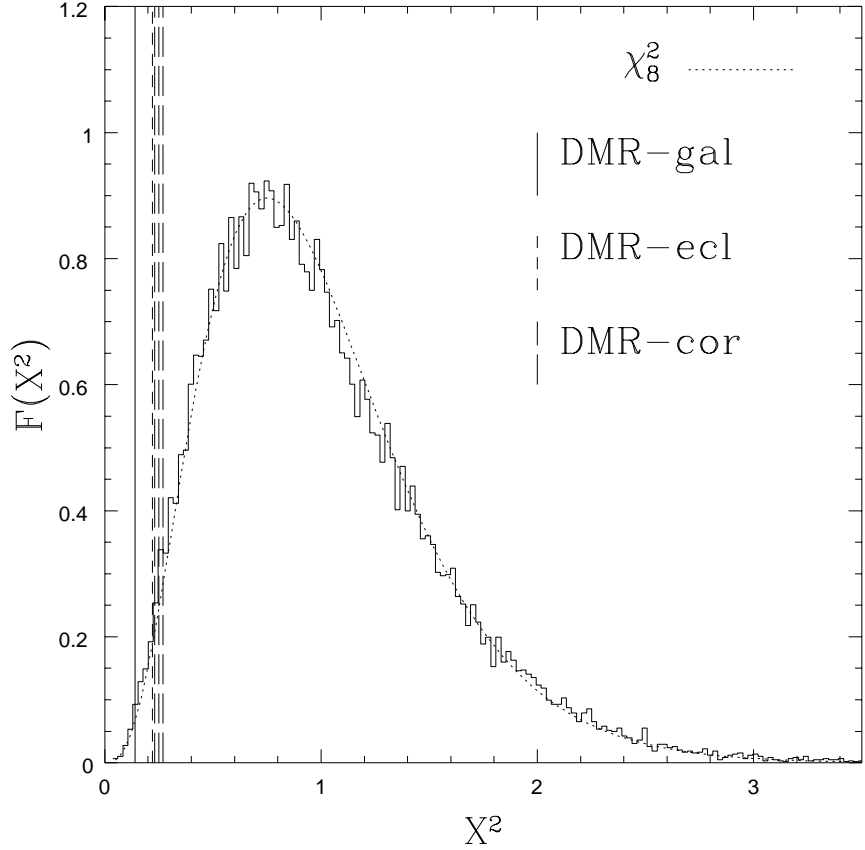,width=12cm}} 
\caption{The distribution $F(X^2)$ as inferred 
from 25000 realizations. 
$F(X^2)$ is well approximated by a $\chi^2_8$. The vertical bars show 
the observed $X^2$ in COBE-DMR 4 year maps in galactic and ecliptic 
pixelizations. We also show the result for the various 
foreground corrected maps - for which there is higher concordance.} 
\label{fig2} 
\end{figure}

We find $X^2=0.14$ and $X^2=0.22$ for data in galactic and ecliptic 
pixelization, respectively. To quantify the confidence level for 
rejecting Gaussianity we determine the distribution of $X^2$,
$F(X^2)$, making use of further 
Monte Carlo simulations. The detailed procedure shadows that described 
in \cite{fmg}. We stress that in each realization a new sky
is produced,  from which a full set of $J_\ell$ is derived, for 
which the $X^2$ is computed. 
The result is plotted in Fig.~\ref{fig2}, where we 
superimpose the observed $X^2$ and its distribution. 
We then compute the  percentage  
of the population with a larger $X^2$ than the observed one.  
We find that $P(X^2>0.14)= 0.998$ (and $P(X^2>0.22)=0.985$) 
for maps in galactic (ecliptic) pixelization.  The lack of scatter 
in the observed $J^3_\ell$ implies that Gaussianity may be rejected 
at the $99.8\%$ ($98.5\%$) confidence level. 
 
A closer analysis reveals that this signal is mainly in the 53 GHz 
channel (see Table 1), which is also the least noisy channel. 
However the confidence level for rejecting Gaussianity increases 
(from $93.2\%$ to $99.8\%$ in galactic pixelization) when the 
53Ghz channel is combined with the 90Ghz channel. Hence the 
overall signal is due to both channels. The reduced confidence 
levels in the separate channels merely 
reflect a lower signal to noise ratio, and the Gaussian nature of 
noise. 
 
\section{Combining different Gaussianity tests} 
Woven into the above argument is a perspective 
on how to combine different Gaussianity tests which
differs from
that presented by \cite{bt99}. In that work the authors argue that 
if $n$ tests are made for a given hypothesis, and they return 
confidence levels for rejection $\{p_i\}$, then, 
if $p_{max}=\max\{p_i\}$, 
the actual confidence level for rejection is $p_{max}^n$. 
 
While the above recipe is formally correct it cannot be applied  
when the hypothesis is Gaussianity. Let the various tests be 
a set of cumulants $\{\kappa_i\}$ (\cite{kos}). Suppose that  
all cumulants are  consistent with Gaussianity except for  
a single cumulant, which prompts us to reject Gaussianity 
with confidence level $p_{max}$.  
Clearly the confidence level for rejecting Gaussianity is $p_{max}$, 
since it is enough for the distribution to have a single non-Gaussian  
cumulant for it to be non-Gaussian. 
 
The point is that Gaussianity cannot be regarded by itself 
as an hypothesis, 
since
 the corresponding alternative hypothesis includes an infinity of 
independent  
degrees of freedom involving different moments and scales. 
The argument of 
\cite{bt99} is correct when applied to independent 
tests concerning the 
same non-Gaussian degree of freedom, for instance independent tests 
related to the skewness. However it cannot be true for different 
tests probing independent non-Gaussian degrees of freedom, say 
skewness and kurtosis.

In the context of our result (which returns 
$X^2\ll 1$),  we  notice that if we were to include into the analysis 
the results of \cite{fmg}  (for which $X^2\gg 1$) 
we would get an average $X^2\approx 1$. Such procedure is obviously
nonsensical: two wrongs don't make a right. One should 
examine independent 
non-Gaussian features separately, in particular the $I^3_\ell$ and the 
$J^3_\ell$, or two ranges of $\ell$, one with
$X^2\gg 1$ the other with $X^2\ll 1$.
 The only practical constraint is 
sample variance, forcing any analysis to include more than one 
degree of freedom so that $F(X^2)$ is sufficiently peaked.

\section{The possibility of a non-cosmological origin} 
Could this signal have a non-cosmological origin?
The possibility of foreground contamination was considered in two
ways. Firstly we subject foreground templates to the same analysis.
At the observing frequencies the obvious 
contaminant should be foreground
dust emission. The DIRBE  maps (\cite{boggess92}) supply us with a
useful template on which we can measure the $J^3_\ell$s. We have done
this for two of the lowest frequency maps, the $100$ $\mu$m  and the
$240$ $\mu$m maps. The estimate is performed in exactly the
same way as for the DMR data (i.e. using the extended Galaxy cut).
We performed a similar exercise with 
the Haslam 408Mhz (\cite{haslam}) map. 
The results are presented in Table 1. We find that the $J^3_\ell$
(in contrast to the $I^3_\ell$ used by \cite{fmg}) are capable of 
exposing the non-Gaussianity in these templates, even
when smoothed by a $7^\circ$ beam. However none of the signatures
found correlates with the DMR signal. DIRBE maps produce highly 
deviant $J^3_\ell$, whereas all $J^3_\ell>0$ for the Haslam map. 

We also considered foreground corrected maps (see Table 1).  In 
these one
corrects the coadded 53 and 90 Ghz maps for the DIRBE 
correlated emission.
We studied maps made in ecliptic and galactic 
frames, and also another map made in the ecliptic frame
 but with the DIRBE
correction forced to have the same coupling 
as determined in the galactic
frame. The confidence levels for rejecting Gaussianity are 
$97.9\%$, $98.5\%$, and $97.1\%$, respectively. The signal is
therefore reduced, but not erased.

Could the observed signal be due to detector noise? The DMR noise
is subtly non-Gaussian,  due to its anisotropy and 
pixel-pixel correlations
(\cite{corn}). These features were incorporated into the simulations 
leading to $P(J^3_\ell)$. However it could just happen that the noise 
in the particular realization we have observed turned out 
to be a fluke,
concentrating the observed $J^3_\ell$ around zero. 
We examined this possibility by considering difference maps 
$(A-B)/2$.
If the observed effect is the result of a noise fluke,
it should be exacerbated in $(A-B)/2$ maps, rather
than in $(A+B)/2$ maps. As can be seen in Table 1, 
one of the noise maps
($(A-B)/2$, 53Ghz, in ecliptic pixelization) is indeed unusually 
non-Gaussian --- but with $X^2\gg 1$ rather than $X^2\ll 1$.  This feature 
disappears in $(A-B)/2$, 53Ghz, 
maps made in the galactic pixelization. 
The $(A-B)/2$, 
90GHz map, in galactic pixelization, has a low $X^2$
but far from significant. 

\cite{bt99} have shown that removing a selected
beam-size region from DMR maps deteriorates the $I^3_\ell$ 
non-Gaussian signal. This fact is of great interest 
as it highlights the possible spatial localization of what
is a priori a ``Fourier space'' statistic. More recently
\cite{banday} pointed out that removing single beam-size 
regions may also reinforce the $I^3_\ell$ 
non-Gaussian signal. We have subjected our $J^3_\ell$ analysis
to this exercise. We found that $X^2$ from maps without a single
beam sized region is {\it very } sharply peaked 
around the uncut value
0.14. There is a region without which $X^2=0.22$ 
but it is also possible to remove a region so that $X^2=0.08$. 
Hence the $J^3_\ell$ signal can never be significantly
deteriorated by means of this prescription, and for this reason
we believe it to be essentially a Fourier space feature.

A number of systematic error templates 
were also examined (\cite{kog96c}).
These provide estimates at the $95\%$ confidence level of errors
due to the following:  the effect of instrument susceptibility 
to the Earth  magnetic field; any unknown effects at the spacecraft 
spin period;
errors in the calibration associated with long-term drifts, and 
calibration errors at the orbit and spin frequency; 
errors due to incorrect removal of the COBE Doppler and  Earth Doppler
signals; errors in correcting for  emissions from the Earth, and
eclipse effects; artifacts due to uncertainty in the correction for
the correlation created by the low-pass filter on the
lock-in amplifiers (LIA) on each radiometer;
errors due to emissions from the moon, and the planets. 
The systematic templates display strongly
non-Gaussian structures, tracing the DMR scanning patterns.
We added or subtracted these templates enhanced by a
factor of up to 4 to DMR maps (see \cite{santa} 
for a better description
of the procedure). The effect on the $J^3_\ell$ spectrum
was always found to be small, leading to very small variations
in the $X^2$.

\cite{banday} have recently claimed that the
systematic errors due to eclipse effects may be larger than
previously thought. They showed how the $I^3_\ell$ change 
dramatically when estimated from maps in which data collected 
in the two month eclipse season has been discarded. 
These maps are more noisy, and the $I^3_\ell$
are very sensitive to noise. Indeed the $I^3_\ell$ are 
cubic statistics, with a signal to noise proportional to
(Number of Observations$)^{3/2}$, and so they are much
more sensitive to noise than the power spectrum.
Perhaps the variations in 
$I^3_\ell$ merely reflect a larger noise, and not a systematic
effect. This possibility could be disproved if no 
striking variations in the $I^3_\ell$ were found in maps
for which {\it other} two month data samples are excised. 
We have applied the $J^3_\ell$ analysis to maps without eclipse
data, and found that the confidence level for rejecting 
Gaussianity does not drop below $99.2\%$ (see Table 1). 
Hence the result described in this Letter appears 
to be robust in this respect.
A more detailed description of the impact of systematics 
upon the $I^3_\ell$ and $J^3_\ell$ will be the subject
of a comprehensive publication (\cite{banday00}).

We finally subject our algorithm to a number of tests. 
Arbitrary rotations 
of the coordinate system (as opposed to the pixelization scheme)
affects $J^3_\ell$ to less than a part in $10^5$. 
Possible residual 
offsets (resulting from the removal of the 
monopole and dipole on the cut
map) do not destroy the signal found. 
Finally changing the various assumptions going into 
Monte Carlo simulations
do not affect the estimated distributions $P(J^3_\ell)$. We found these
distributions to be independent of the assumed shape of 
the power spectrum,
of the exact shape of the DMR beam, or the inclusion of the pixel
window function.

In summary the $J^3_\ell$ analysis appears to be more sensitive
to shortcomings in DMR maps than the $I^3_\ell$. This fact is
already obvious in the differences between ecliptic and galactic
pixelizations in the publicly released maps. When all possible 
renditions of DMR data are considered 
the significance 
level of our detection may vary between $97\%$ and $99.8\%$.
Therefore the various tests for systematics
we have described do not leave the result 
unscathed, but neither do they rule out a cosmological origin.
We should stress that, in line with all previous work in the field,
we have employed a frequentist approach. One may therefore question
the Bayesian meaning of the confidence levels quoted. 
A Bayesian treatment of the bispectrum remains  unfeasible
(see however \cite{contaldi}).

In comparison with \cite{fmg} the result we have described is 
more believable from a theoretical point of view. It spreads 
over a range of scales. Previous detections  concentrate 
on a single mode. The result obtained is puzzling in that,
rather than revealing the presence of deviants, it shows a perfect 
alignment of the observed $J^3_\ell$ on
 the top of their distribution
for a  Gaussian process. This is perhaps not as 
strange as it might seem at 
first: in \cite{fms} it was shown how 
non-Gaussianity may reveal itself not by 
non-zero average cumulants, but by 
abnormal errorbars around zero. 
 
 
\section*{ACKNOWLEDGEMENTS} 
I would like to thank K. Baskerville, P. Ferreira and 
K. Gorski for help with this project. Special thanks to T. Banday
for supplying DMR 
maps without the eclipse season data. This work was performed 
on COSMOS, the Origin 2000 supercomputer owned by the UK-CCC and 
supported by HEFCE and PPARC. I thank the Royal Society 
for financial support.

\begin{table} 
\label{table1} 
\begin{center} 
\begin{tabular}{cccccccccccc} 
\tableline 
  & $J^3_4$ & $J^3_6$ & $J^3_8$ & $J^3_{10}$ & $J^3_{12}$ & 
 $J^3_{14}$ & $J^3_{16}$ & $J^3_{18}$ & $X^2$ & Reject \%\\ 
\tableline 
\tableline Gauss-rms&.263&.225&.194&.178&.162&.153&.144&.135& 
---&---\\ 
\tableline
DMR - ecl& -.521& .009 & -.022 &-.007 &.112 & .088&  .054 & 
.045 & .22 & 98.5 \\  
DMR - gal&-.502&.012&.029&-.088&.098&-.002&.013&-.030&.14& 
 99.8\\ 
DMR - cor/ecl & -.554& .022 & -.091 &-.099 &.104 & .061&  .071 & 
.057 & .25 & 97.9 \\
DMR - cor/ecg & -.555& .026 & -.105 &-.104 &.102 & .055&  .076 & 
.059 & .27 & 97.1 \\
DMR - cor/gal&-.542&.042&-.068&-.140&.108&-.026&.031&-.019&.23& 98.5\\ 
\tableline
DMR - gal 90&-.513&.003&-.316&-.167&.140&.063&.042&-.049&.66& 
 71.2\\ 
DMR - gal 53&-.497&-.053&-.009&.022&.172&-.076&.107&-.031&.36& 
 93.2\\ 
\tableline
DMR - gal/ne &-.448&.034&.042&-.022&.127&.009&.008&.042&.18& 
99.2 \\
\tableline
A-B 53 ecl &-.304&-.002&.126&-.008&-.426&.437&.170&-.047&2.41
& 98.4 $^*$\\ 
A-B 53 gal &-.259&.034&.118&-.003&-.193&.248&.246&.019&1.07& 38.1\\ 
A-B 90 ecl &-.134&-.151&-.195&.099&-.181&.091&-.273&.021&1.00& 43.3\\ 
A-B 90 gal &-.085&-.158&-.173&.001&-.140&.047&-.180&.079&.57& 79.0\\ 
\tableline
DIRBE08 ecl &.076&.203&-.109&.404&.240&.027&-.293&-.107&1.93
& 94.1$^*$\\ 
DIRBE10 ecl &.252&.124&-.267&.706&.134&.184&-.177&-.145&3.15
&99.7$^*$\\ 
Haslam ecl&.279&.459&.062&.158&.196&.146&.049&.100&1.21&29.4\\ 
Haslam gal &.258&.462&.044&.182&.163&.105&.037&.045&1.04&39.7\\ 
\tableline 
\end{tabular} 
\end{center} 
\caption{The values of $J^3_\ell$ for various datasets, 
their $X^2$, and the confidence level for rejecting Gaussianity 
on the grounds of $X^2\ll 1$ (starred ($^*$) figures indicate 
confidence levels for rejecting Gaussianity on the grounds of 
$X^2\gg 1$). ``ecl'' and ``gal'' 
stand for ecliptic and galactic pixelizations, ``cor'' for foreground
corrected, ``ne'' for no-eclipse data. 
On the first line we show the rms for a Gaussian process.
This table is more illuminating than a plot, as most $J^3_\ell$
accumulate around zero.} 
\end{table}

\end{document}